\documentclass{article}
\usepackage{frascatiphys}
\usepackage{graphicx}
\begin{document}
\title{ 
GAMMA RAYS FROM CLUSTERS OF GALAXIES
}
\author{
Stefano Gabici        \\
{\em Dipartimento di Astronomia e Scienza dello Spazio,}\\{\em Largo E. Fermi 5, 50125 Firenze, Italy} \\{\em gabici@arcetri.astro.it}\\
}
\maketitle
\baselineskip=11.6pt
\begin{abstract}
The non thermal radiation observed from a handful of clusters of galaxies (CG) is the proof that particle acceleration occurs in the intra cluster medium (ICM). It is often believed that shock surfaces associated with either mergers of CG, or with the cosmological inflow of matter onto clusters during structure formation may be the sites for acceleration. We discuss here the effectiveness of shock acceleration in the ICM, stressing that merger related shocks are typically weak, at least for the so-called major mergers. We investigate the implications of shock strengths for gamma ray emission from single CG and for their detectability with future gamma ray satellites (such as GLAST and AGILE) and ground based Cherenkov telescopes. We also discuss the contribution of clusters to the extragalactic diffuse gamma ray background (EDGRB).
\end{abstract}
\baselineskip=14pt
\section{Introduction}

Clusters of galaxies are powerful X-ray sources, with luminosities in the range $10^{43} - 10^{46} erg \, s^{-1}$. This thermal radiation is emitted by an hot ($\sim 10^8 K$), tenuous ($\sim 10^{-3} cm^{-3}$), highly ionized intra cluster gas that fills all the cluster volume and accounts for a dominant fraction of the cluster baryons \cite{sarazin}.
Compelling evidence for the presence of relativistic particles and magnetic field in the intra cluster medium (hereafter ICM) is given by the observation of a diffuse synchrotron radio emission in a consistent fraction of the rich clusters \cite{gigia}. This gives us a direct proof of the presence of GeV electrons and $\sim \mu G$ magnetic field. The same electrons responsible for the radio emission can upscatter the cosmic microwave background (CMB) photons in the X-ray band via inverse Compton scattering (ICS) providing an explanation for the hard X-ray excess observed in a few clusters \cite{fusco,jfb}. Different mechanisms, such as non thermal Bremsstrahlung from a population of supra thermal electrons, have also been proposed \cite{stoc}. 

Clusters are also expected to be potential gamma ray sources, the main emission processes being inverse Compton scattering off the cosmic microwave background photons (CMB) from ultra relativistic electrons and the decay of neutral pions produced in the interactions between cosmic ray protons and ICM protons \cite{pasquale}.
In this paper we make predictions on the number of clusters that future gamma ray telescopes (both from space and earth) will be able to detect and on the contribution of clusters to the extragalactic diffuse gamma ray background. We focus our attention on the photons produced by electrons accelerated at shocks that develop during the process of large scale structure formation. 
Protons are also accelerated at shocks, they cannot escape the cluster volume \cite{venia} and they are re-energized each time a new merger event occurs \cite{io1}. It follows that the spectrum and the energy density of protons depend on the merger history of the cluster and an hadronic feature might appear on top of the gamma ray emission generated by inverse Compton scattering of electrons\cite{dermer}.
 
\section{Shock acceleration during structure formation}

The ICM is heated to the observed high temperature by shock waves that naturally form during the process of large scale structure formation \cite{ryu}. Diffusive acceleration can take place at these shocks so that a small fraction of the particles crossing the shock discontinuity can be energized up to relativistic energies. The maximum energy for the accelerated particles can be evaluated equating the acceleration time with the energy loss time. 
Under the reasonable assumption that during the acceleration process the residence time of the particle in the upstream region is longer than the residence time downstream, and neglecting a weak dependence on the shock Mach number we can approximate the acceleration time as \cite{io3}:
\begin{equation}
\tau_{acc}(E) \sim \frac{4D(E)}{u^2}
\end{equation}
where $D$ and $u$ are the upstream diffusion coefficient and the upstream fluid velocity. Under these hypothesis the acceleration time depends only on upstream quantities, unaffected by the presence of the shock.
Assuming a Bohm diffusion coefficient we obtain:
\begin{equation}
\label{eq.acc}
\tau_{acc}(E) \sim 0.3 B(\mu G)^{-1} E(GeV) (\frac{u}{10^8 cm s^{-1}}) \; years.
\end{equation}
For electrons the relevant loss channel is ICS on the CMB radiation characterized by a time scale:
\begin{equation}
\label{eq.loss}
\tau_{ICS} = \frac{E}{\frac{4}{3}\sigma_TcU_{CMB}\gamma^2} \sim 10^9 E(GeV)^{-1}(1+z)^{-4} \; years
\end{equation}
where $U_{CMB}$ is the CMB energy density and $\gamma$ is the electron Lorentz factor.
Equating \ref{eq.acc} and \ref{eq.loss} we obtain the maximum energy for the electrons:
\begin{equation}
E_{max} \sim 57 B(\mu G)^{1/2} (\frac{u}{10^8 cm s^{-1}}) \; TeV
\end{equation}
corresponding to a maximum energy for the upscattered CMB photons equal to:
\begin{equation}
\label{eq.max}
E_{\gamma}^{max} = \frac{4}{3}\gamma^2 \epsilon_{CMB} \sim 7.5 B(\mu G) (\frac{u}{10^8 cm s^{-1}})^2 \; TeV
\end{equation}
which falls in the very high energy gamma ray band.
It is important to stress, however, that the Bohm diffusion coefficient is the smallest possible one, and that a different choice of the diffusion coefficient could result in a too low maximum energy of the electrons, so that the electron population would be unable to upscatter the CMB photons up to gamma ray energies.
In the following we use for the maximum energy the value obtained in eq. \ref{eq.max}.

The spectra of particles accelerated at shocks are power laws in momentum of the form $N(p) \propto p^{-\alpha}$, where the slope $\alpha$ is connected to the shock Mach number ${\mathcal M}$ through the well known relation:
\begin{equation}
\alpha = 2 \, \frac{{\cal M}^2+1}{{\cal M}^2-1} .
\end{equation}

\section{Large scale structure formation and associated shocks.}

\begin{figure}[htb]
  \begin{center}
    \includegraphics[angle=0,
width=0.5\textwidth]{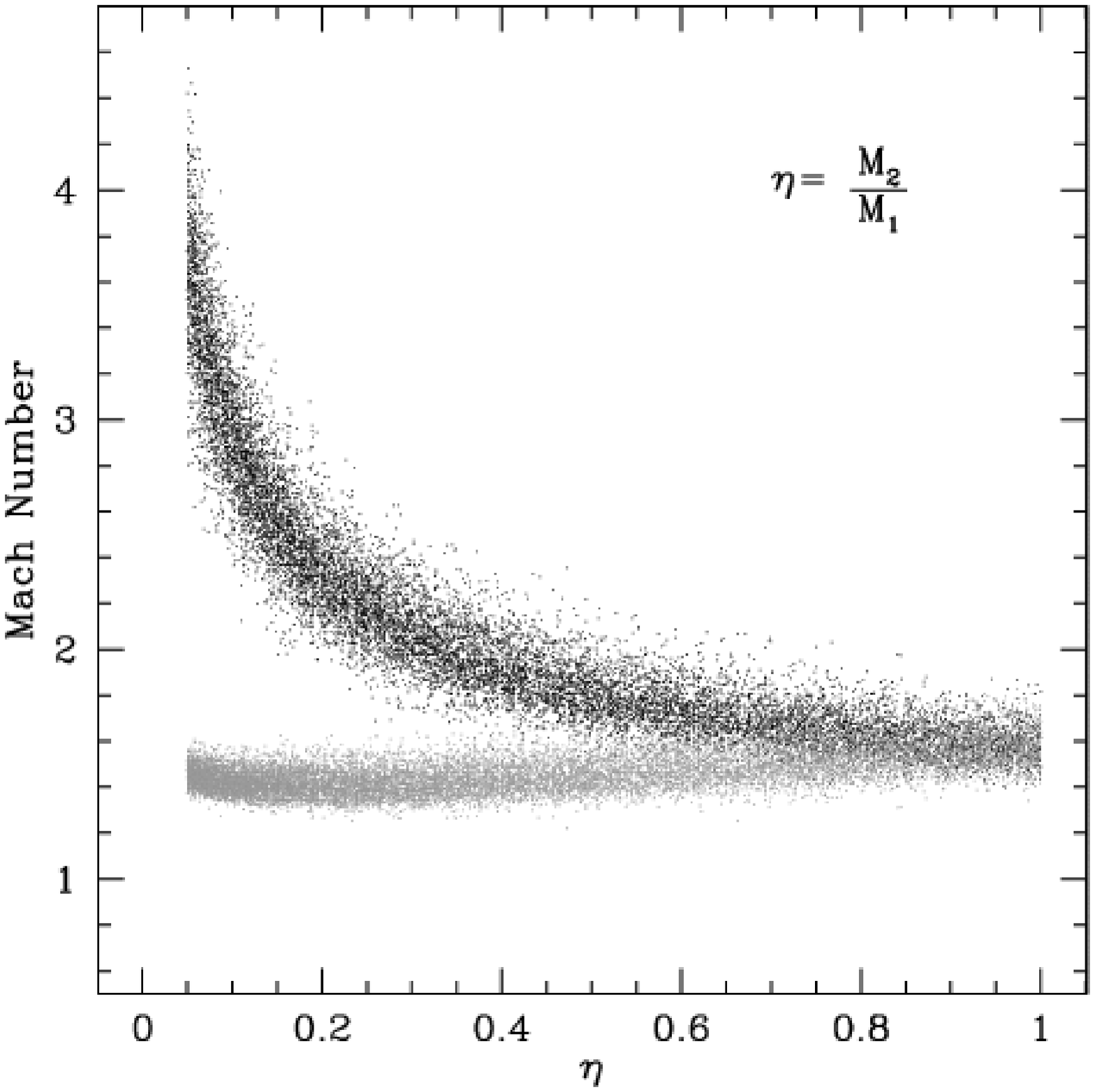}
    \includegraphics[angle=0,
width=0.48\textwidth]{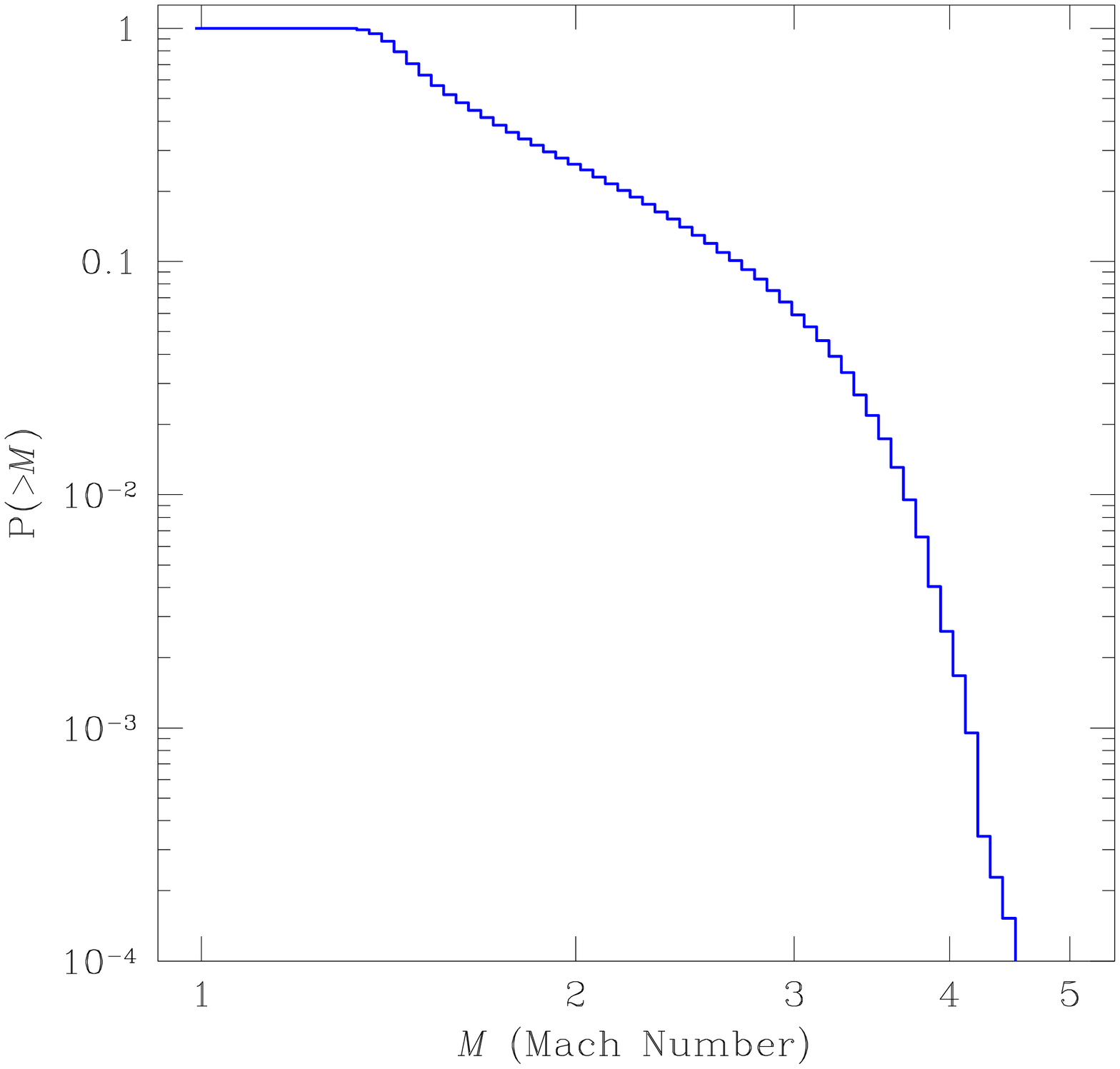}
    \caption{Merger shock Mach numbers as a function of the ratio between the masses of the two colliding clusters (left panel) and Mach number cumulative distribution (right panel). The upper strip in the left panel refers to the shocks that propagate in the smaller merging cluster while the lower strip refers to the shocks propagating in the bigger, hotter cluster.}
    \label{mach}
  \end{center}
\end{figure}

According to the standard theory of large scale structure formation clusters form hierarchically through mergers of smaller subunits. Semi-analytical models describing this process can be found in the literature \cite{PS,LC}. This formalism provides us the comoving density of objects with given mass $M$ at a redshift $z$, $n(M,z)$, and the rate at which clusters of mass $M_1$ and $M_2$ merge together at a fixed redshift $z$, $R(M_1,M_2,z)$. Sampling these distributions it is possible to simulate the whole merger history of a cluster with a certain present mass.
Each time a merger occurs we can estimate the relative velocity $v_r$ between the two involved clusters using energy conservation \cite{taki,io1}:
\begin{equation}
-\frac{GM_1M_2}{R_1+R_2}+\frac{1}{2}M_rv_r^2=-\frac{GM_1M_2}{2R_{ta}}
\end{equation}
where $M_i$ and $R_i$ are the masses and virial radii of the two colliding clusters, $M_r$ is the reduced mass and $R_{ta}$ is the turnaround radius of the system.
The shock Mach number is the ratio between $v_r$ and the sound speed in the ICM, given by the virial theorem:
\begin{equation}
c_{s,i}^2 = \gamma(\gamma-1)\frac{GM_i}{2R_i}.
\end{equation}
In figure \ref{mach} (left panel) the Mach numbers $\cal M$ of the shocks are plotted as a function of the ratio between the masses of the two merging clusters. This result has been obtained simulating 500 different realizations of a merger tree for a cluster with present mass $10^{15}$ solar masses \cite{io1}. In the right panel the cumulative distribution of Mach numbers is shown.

Fig. \ref{mach} tells us that merger shocks are often weak, following a distribution strongly peaked at ${\cal M} \sim 1.4$ and consequently are unable to accelerate relativistic particles in an efficient way (accelerated spectra are very steep power law with $\alpha \sim 6$). Only $\sim 6\%$ of the shocks have Mach number greater than 3 (corresponding to $\alpha > 2.4$), necessary to be consistent with the spectral shape of cluster radio halos, and these strong shocks form only during minor mergers (collisions between clusters with very different masses). On the contrary radio halos are found in a consistent fraction ($\sim 40\%$) of massive clusters and there seems to be a correlation between the presence of the halo and the signature of a recent major merger \cite{gigia}. We can conclude that radio halos are unlikely to be produced by electrons directly accelerated at merger shocks \cite{io1}.

Besides merger shocks, that develop in the ICM of virialized objects, we consider here also accretion shocks that form due to the secondary infall of non virialized matter onto an already formed cluster \cite{bert}. These shocks propagate in a cold medium and are strong by definition, so that the spectra of the accelerated particles are fixed to $N(p) \propto p^{-2}$. We assume here that a spherical shock forms at the virial radius of each cluster and that the cold infalling matter (whose density is taken here equal to the average baryon density of the universe) flows at the free fall velocity.

\section{Source counts and the contribution to the gamma ray background}

Recently an association between clusters of galaxies and unidentified EGRET sources was proposed by different authors \cite{cola,gll}. The physical plausibility and the statistical significance of such an association were strongly questioned in \cite{reimer}, and the lack of association was also found in \cite{sm}.
Future gamma ray space telescopes, such as AGILE and GLAST, will observe the whole sky with a better sensitivity compared to EGRET. We make here predictions on the number of objects that these telescopes will be able to detect.

The number of accreting clusters with gamma ray flux greater than a fixed value $F_{lim}$ can be evaluated as follows:
\begin{equation}
N(>F_{lim})=\int_0^{\infty} dz\,\frac{dV}{dz} \int_{M(F_{lim},z)}^{\infty} dM \, n(M,z)
\end{equation}
where $dV$ is the comoving volume between $z$ and $z+dz$, $n(M,z)$ is given in \cite{PS}, and $M(F_{lim},z)$ is the mass of a cluster accreting at redshift $z$ whose flux is $F_{lim}$.
For merging clusters a similar, slightly more complicated expression, which includes also the merger rate $R(M_1,M_2,z)$, holds.
Gamma ray luminosities are evaluated assuming that a small fraction $\eta \sim 5\%$ of the kinetic energy flowing across the shock is converted in relativistic electrons, which upscatter the CMB photons in the gamma ray energy band.

\begin{figure}[htbp]
  \begin{center}
    \includegraphics[angle=0,
width=0.48\textwidth]{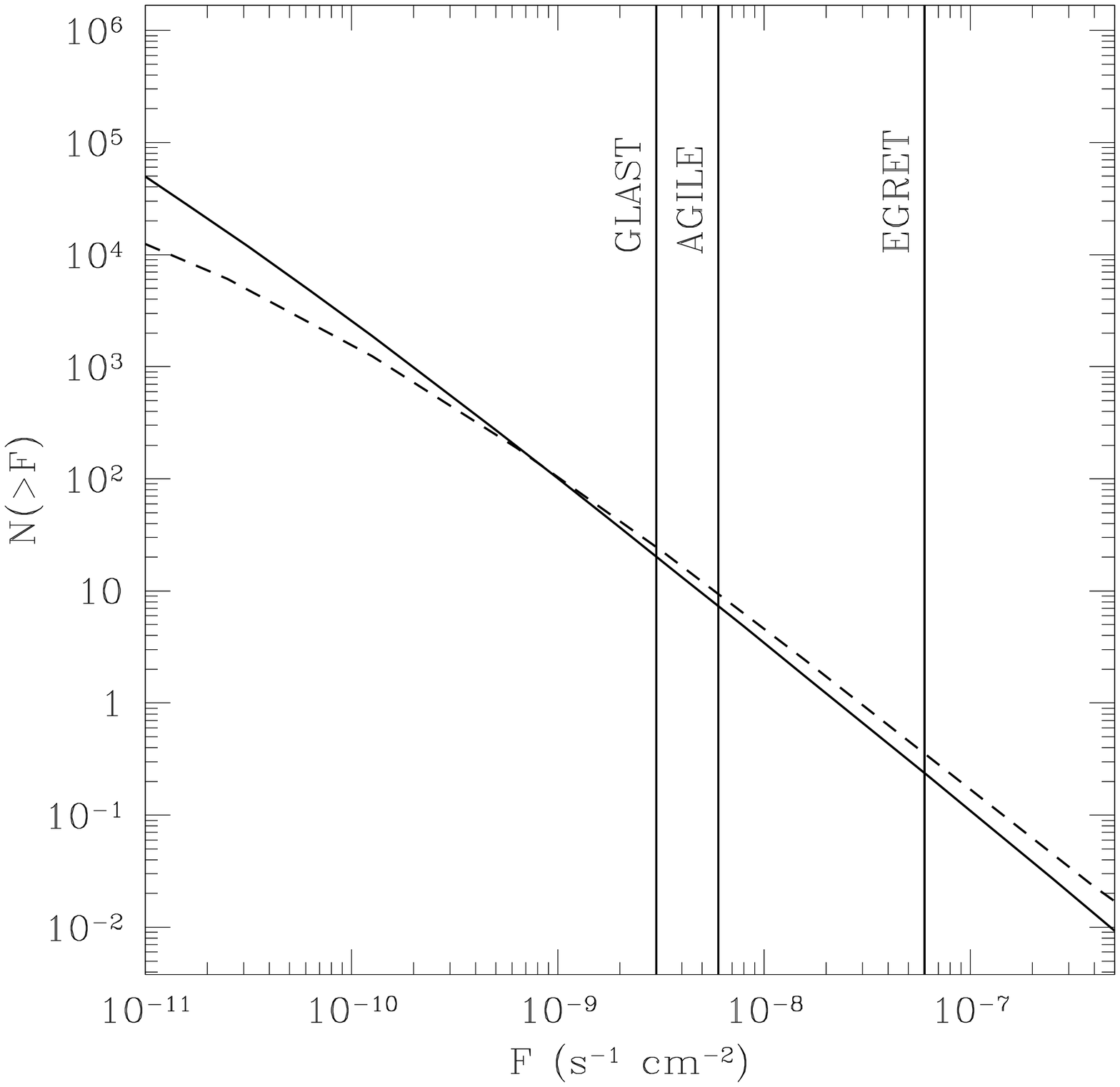}
    \includegraphics[angle=0,
width=0.5\textwidth]{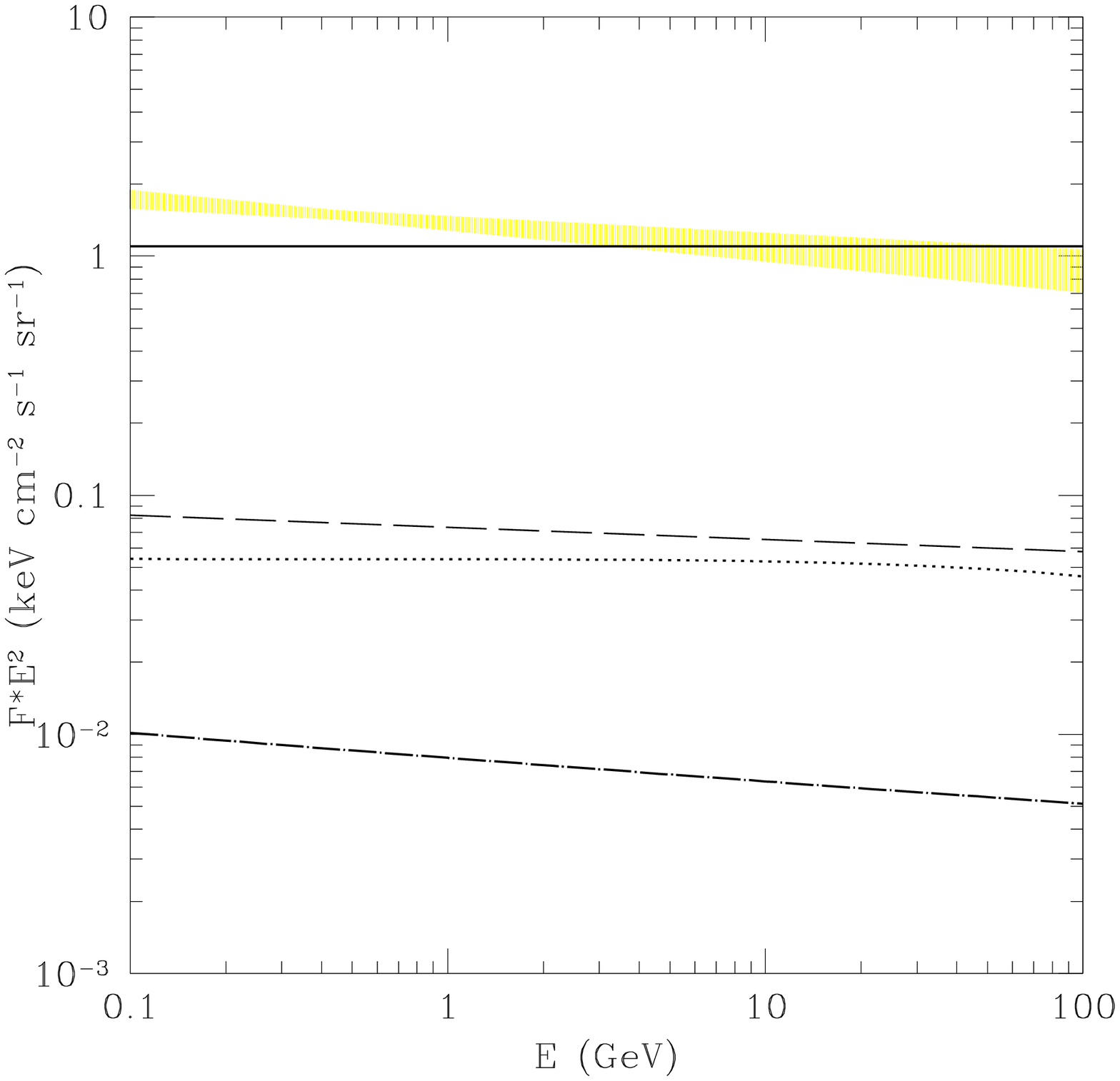}
    \caption{Number of accreting (solid line) and merging (dashed line) clusters with gamma ray flux greater than $F$ (left panel). In the right panel are shown our predictions for the extragalactic diffuse gamma ray background (see text for details).}
    \label{gamma}
  \end{center}
\end{figure}

The results, as obtained in \cite{io3}, are shown in fig \ref{gamma} (left panel) and can be summarized in this way: AGILE and GLAST will be able to observe $\sim 10$ and a few tens of clusters respectively (both merging and accreting), while no one of the unidentified EGRET sources is expected to be associated with a cluster, in agreement with \cite{reimer}.

The superposition of gamma ray emission from single clusters results in a diffuse background. In \cite{wax} was firt proposed that the CMB photons, scattered up to gamma ray energies by relativistic electrons accelerated at the shocks that form during large scale structure formation, could account for the whole extragalactic diffuse gamma ray background, as estimated subtracting a galactic emission model to the EGET data \cite{egret}.
We evaluate in \cite{io2} this contribution using the formalism described in the previous sections. Our results are shown in fig. \ref{gamma} (right panel). The shaded region corresponds to the EGRET data and the solid line that roughly fits the data is the prediction made in \cite{wax}. Our estimates of the contribution to the background from accreting and merging clusters are represented by the dotted and dashed lines respectively. The dot-dashed line represents a more realistic estimate for the diffuse emission due to merging clusters, obtained considering only objects with mass greater than $10^{13} M_{\odot}$ (the typical mass of a group of galaxies).
According to our estimate, forming structures in the universe can account only for $\sim 10\%$ of the extragalactic background and this is in agreement with recent results from detailed numerical simulations \cite{uri,francesco}.

\section{Very high energy gamma ray emission}

\begin{figure}[htb]
  \begin{center}
    \includegraphics[angle=0,
width=0.7\textwidth]{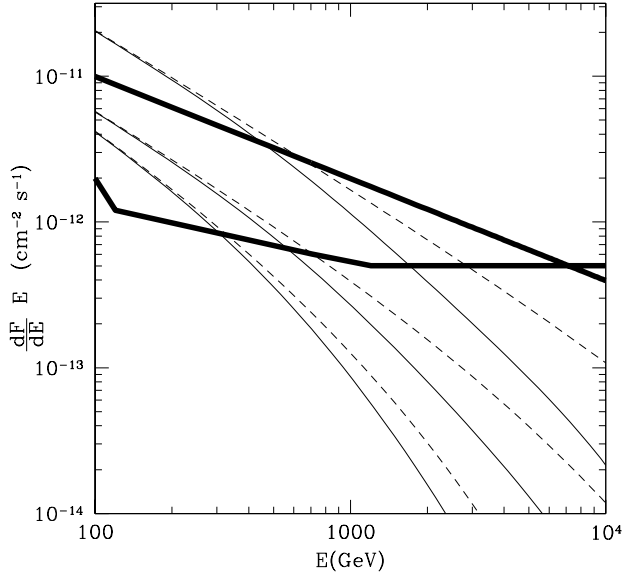}
    \caption{Gamma ray emission in the 100 GeV - 10 TeV region. The thick solid lines represent the sensitivities of a cherenkov array for point sources (lower curve) and extended sources (upper curve). The predicted gamma ray fluxes from a Coma-like cluster with and without absorption of the infrared background are plotted as dashed and solid lines respectively.}
    \label{TeV}
  \end{center}
\end{figure}

The maximum energy of the upscattered CMB photons falls in the TeV range (see eq.\ref{eq.max}) so that we can wonder if clusters could be detected by ground based Cherenkov telescopes. In fig. \ref{TeV} we plot the sensitivities for a generic telescope as calculated in \cite{felix} for 100 hours of observation with an array consisting of 10 cells (pointlike and $1^{\circ}$ sources) together with the expected spectra for three different cases (top to bottom): 1) a merger between two clusters with masses $10^{15}$ and $10^{13} M_{\odot}$, 2) a $10^{15} M_{\odot}$ accreting cluster with a magnetic field in the upstream region equal to $0.1 \mu G$, 3) the same as 2) but with a ten times lower magnetic field. All the considered clusters are at a distance of $100 Mpc$. The thin continuous and dotted lines refer to the spectra calculated with and without taking into account the absorption on the infrared background \cite{stecker}.

The apparent size of a $10^{15} M_{\odot}$ accreting cluster at a distance of $100 Mpc$ is comparable or even greater than the field of view of Cherenkov telescopes ($\sim$ a few degrees), while the size of the emitting region for the merger considered in fig. \ref{TeV} should be roughly $1^{\circ}$, making it marginally detectable.

\section{Conclusions}

In \cite{io1} we first proposed that merger shocks are often weak and consequently unable to accelerate efficiently relativistic particles. We developed a recipe to evaluate the shock Mach numbers as a function of the masses of the two merging objects finding that shock acceleration can be effective only during minor mergers. We extended our analysis in \cite{io2} where we included also the effects of strong accretion shocks. 

Ultrarelativistic electrons can be accelerated at such strong shocks and they can upscatter the CMB photons to very high gamma ray energies, making clusters of galaxies potential GeV and TeV gamma ray sources. We estimate that the future gamma ray telescopes AGILE and GLAST will observe $\sim 10$ and $\sim 50$ clusters respectively \cite{io3}, while we do not expect any association between clusters and unidentified EGRET sources, in agreement with the observational results presented in \cite{reimer}. We also estimated the contribution from large scale structure formation to the EDGRB to be at most equal to $\sim 10\%$ of the observed one.

Only massive, nearby merging clusters could be detected by future Cherenkov arrays, while the detection of accreting clusters seems to be more problematic due to the very extended apparent size of these sources.

\section{Acknowledgments}
I am grateful to G. Brunetti for useful comments.

\end{document}